\documentclass{pasj00}
\newcommand{\solarmass}{\mbox{${\rm M_{\odot}}$}}

\def\ltsima{$\; \buildrel < \over \sim \;$}
\def\simlt{\lower.5ex\hbox{\ltsima}}
\def\gtsima{$\; \buildrel > \over \sim \;$}
\def\simgt{\lower.5ex\hbox{\gtsima}}

\begin{document}
\SetRunningHead{Kubota et al.}{Suzaku and Optical Spectroscopic Observations of SS~433}
\Received{2009/11/11}%{yyyy/mm/dd}
\Accepted{2010/01/15}%{yyyy/mm/dd}

\title{Suzaku and Optical Spectroscopic Observations of SS~433 in the
2006 April Multiwavelength Campaign}

%%% begin:list of authors
% Do NOT capitalize all letters in "textsc".
\author{Kaori \textsc{Kubota},\altaffilmark{1}
Yoshihiro \textsc{Ueda},\altaffilmark{1}
Nobuyuki \textsc{Kawai},\altaffilmark{2}
Taro \textsc{Kotani},\altaffilmark{3}
Masaki \textsc{Namiki},\altaffilmark{4}
Kenzo \textsc{Kinugasa},\altaffilmark{5}
Shinobu \textsc{Ozaki},\altaffilmark{6}
Takashi \textsc{Iijima},\altaffilmark{7}
Sergei \textsc{Fabrika},\altaffilmark{8}
Takayuki \textsc{Yuasa},\altaffilmark{9}
Shin'ya \textsc{Yamada},\altaffilmark{9}
Kazuo \textsc{Makishima},\altaffilmark{9}
}
\altaffiltext{1}{Department of Astronomy, Kyoto University, Sakyo-ku,
Kyoto 606-8502}
\email{kaori.k@kusastro.kyoto-u.ac.jp}
\altaffiltext{2}{Tokyo Institute of Technology}
\altaffiltext{3}{Aoyama Gakuin University}
\altaffiltext{4}{Osaka University}
\altaffiltext{5}{Gunma Astronomical Observatory}
\altaffiltext{6}{Okayama Astrophysical Observatory, 
National Astronomical Observatory of Japan}
\altaffiltext{7}{Astronomical Observatory of Padova}
\altaffiltext{8}{Special Astrophysical Observatory}
\altaffiltext{9}{Department of Physics, University of Tokyo}
%%% end:list of authors

%% `\KeyWords{}' always has to be placed before `\maketitle'.
%Do NOT move this preamble from here!

\maketitle

\begin{abstract}

We report results of the 2006 April multi-wavelengths campaign of
SS~433, focusing on X-ray data observed with Suzaku at two orbital
phases (in- and out-of- eclipse) and simultaneous optical
spectroscopic observations. By analyzing the Fe$_{\rm XXV}$ K$\alpha$
lines originating from the jets, we detect rapid variability of the
Doppler shifts, ${\rm d}z/{\rm d}t \approx 0.019/0.33$ day$^{-1}$,
which is larger than those expected from the precession and/or nodding
motion. This phenomenon probably corresponding to ``jitter'' motions
observed for the first time in X-rays, for which significant
variability both in the jet angle and intrinsic speed is
required. From the time lag of optical Doppler curves from those of
X-rays, we estimate the distance of the optical jets from the base to
be $\approx (3-4)\times 10^{14}$ cm. Based on the radiatively
cooling jet model, we determine the innermost temperature of the jets
to be $T_0$ = 13$\pm$2 keV and 16$\pm$3 keV (the average of the blue
and red jets) for the out-of-eclipse and in-eclipse phase,
respectively, from the line intensity ratio of Fe$_{\rm XXV}$
K$\alpha$ and Fe$_{\rm XXVI}$ K$\alpha$. While the broad band
continuum spectra over the 5--40 keV band in eclipse is consistent
with a multi-temperature bremsstrahlung emission expected from the
jets, and its reflection component from cold matter, the
out-of-eclipse spectrum is harder than the jet emission with the base
temperature determined above, implying the presence of an additional
hard component.

\end{abstract}

\section{Introduction}

SS~433 is a unique Galactic binary system that exhibits continuous
relativistic jets, and hence is an ideal target to study fundamental
problems of astrophysical jets, including the acceleration/collimation
mechanism, structure, and influence on the environment (for a review,
refer to e.g., \cite{margon1984,fabrika2004}). The intrinsic speed of
the jets is 0.26$c$ ($c$ is the light speed), and the jet axis
precesses with a period of 162~days. The optical and X-ray spectra show
several pairs of Doppler shifted lines from the bipolar jets. The
strong emission lines, such as H$\alpha$ and lines from ionized heavy
elements, give direct evidence that the jets mainly contain baryonic
plasmas. Most recently, \citet{kubota2009} constrains the mass of the
compact object to be 1.9--4.9 \solarmass, indicating that the
supercritical accretion star is most likely a low mass black hole,
although a possibility of a massive neutron star cannot be ruled
out. The inclination angle and orbital period are accurately measured
to be $i=78.8^\circ$ \citep{margon1989} and $P$=13.082 days
\citep{goranskii1998}, respectively.

Soon after the discovery of SS~433, many efforts were done to determine
the kinematic model and parameters of the jets, by tracing the Doppler
shifts of emission lines in the optical spectra.  \citet{margon1979} for
the first time proposed a kinetic model, where the jets of about 1/4 the
light speed make precession, with a period of 162.15 days, according to
the latest value obtained by \citet{gies2002}. In superposition to the
``precession'' motion, another regular periodicity of the Doppler shifts
is observed as a ``nodding'' motions with an amplitude of $\Delta
z \sim 0.01$ and a period of 6.28 days. They are caused by the tidal
torques from the companion star.  Furthermore, faster random variability
of Doppler shifts were sometimes reported (see e.g.,
\cite{ciatti1981,katz1982,iijima1993,collins1994}). These phenomena are
called as ``jitter'', which contains an important clue to understand the
collimation mechanism of the jets. The origins of ``jitter'' have not
been understood, in particular whether they are caused by the change of
the speed or orientation due to insufficient observational quality
\citep{collins1994}.

SS~433 is also a target of great interest in high energy astrophysics,
and has been observed by many satellites from early on (e.g., with
EXOSAT, \cite{watson1986}; Tenma, \cite{matsuoka1986}; Ginga,
\cite{kawai1989}). The ASCA satellite, which carried X-ray CCD cameras
for the first time, detected many pairs of Doppler shifted emission
lines from ionized metals such as Si, S, Ar, Ca, Fe and Ni
originating from the twin jets \citep{kotani1994}. By detailed
analysis of the line intensities of Fe$_{\rm XXVI}$ K$\alpha$ to
Fe$_{\rm XXV}$ K$\alpha$, \citet{kotani1996} estimated the length of
the X-ray jets and temperature of the jet base to be $\sim$ 10$^{13}$
cm and $\sim$20 keV, respectively. The emission lines are resolved
with the Chandra HETGS, which are found to have Doppler widths of
$\approx$1000--5000 km s$^{-1}$
\citep{marshall2002,namiki2003,lopez2006}. The
broad band continuum (up to 100 keV) is approximated by a thermal
bremsstrahlung spectrum with a temperature of $\sim$10--30 keV,
depending whether in- or out-of-eclipse 
\citep{kawai1989,cherepashchuk2005}. Basically, hydrodynamical
radiative plasma flow models are found to successfully apply to the
data \citep{brinkmann1991,kotani1996}. Additional complex features are
detected from the XMM spectra, however, which could be Compton
scattered emission from the jet base \citep{brinkmann2005} or an
iron-K absorption edge due to partial covering
\citep{kubota2007}. From the width of an eclipse in the 25--50 keV
band with INTEGRAL, \citet{cherepashchuk2007} and
\citet{krivosheyev2009} propose that a hot extended corona around the
accretion disk is responsible for the hard X-ray emission via thermal
Comptonization with a temperature of $\sim$20 keV. High quality X-ray
spectra covering the broad band are critical to establish the
interpretation of the high energy spectra of SS~433.

Simultaneous, multi-wavelength observations of this source are
particularly important to reveal the structure of the jets. Previous
studies revealed that X-rays are emitted from a region close to the
base of the jets ($\sim 10^{10-13}$ cm; \cite{kotani1996}), while the
optical emission originates far away from the jet base at a distance
of $\sim 10^{14-15}$ cm \citep{fabrika2004}. Radio synchrotron emission
is observed on a larger scale ($\ge 10^{15}$ cm). Comparison of
(quasi) simultaneous data enables us to trace the evolution of the
same matter traveling along the jets on a spatial scale of many orders
of magnitudes.

For this purpose, we organized a multi-wavelength observation campaign
in the period from 2006 March 29 to April 15. The campaign includes
two Suzaku observations performed in- and out-of eclipse phase and a
large set of optical spectroscopy, optical/IR photometry and radio
observations. As for details of this campaign, refer to the papers by
Kotani et al. (2006, 2008).
%\citet{kotani2006,kotani2008}. 
In this paper, we present the results
from the Suzaku observations and those from optical
spectroscopy. Suzaku, the 5th Japanese X-ray satellite, covers the
wide energy band from 0.2--600 keV. The large effective area below 10
keV with good energy resolution, and high sensitivity above 10 keV,
provides us the best opportunity to investigate time variability of
the wide band spectrum of SS~433.  \S~2 describes the observations
and data reduction. The results based on the Suzaku data of
out-of-eclipse are summarized in \S~3, and those of in-eclipse are in
\S~4. We discuss our results in \S~5, including the comparison of
the X-ray and optical data.

\section{Observations and Data Reduction}

\subsection{Suzaku}

We observed SS~433 with Suzaku on two occasions, on 2006 April 4--5
and April 8--9 for a net exposure of $\approx$40 ks each. Suzaku
\citep{mitsuda2007} carries the four X-ray Imaging Spectrometers
(XIS-0, 1, 2 and 3) coupled with the X-ray Telescopes (XRTs) and the
Hard X-ray Detector (HXD), which covers the 0.2--10 keV and 10--600
keV bands, respectively. The XIS-0, 2, and 3 are front-side
illuminated CCDs (XIS-FI), while the XIS-1 is a back-side illuminated
one (XIS-BI).  The energy resolution of the XISs is $\approx$130 eV
(FWHM) at 6 keV. The first observation was performed at the orbital
phase of $\phi=0.97$ (as calculated from the ephemeris by
\cite{goranskii1998}), during the midst of an eclipse by the companion
star, while the second was at $\phi=0.26$ after the source recovered
from the eclipse. From our optical observations, the precession phase
is found to be $\psi \simeq 0.6$ (as calculated from the ephemeris by
\cite{gies2002}), when we see the accretion disk in almost an edge-on
view.  
 At this precession phase, the directions of the twin jets with respect
to the line-of-sight are inverted from those observed at most of the
precession phases, and both show redshifts. In this paper, we call the
jet with a larger redshift as ``red'' jet and the other as ``blue''
jet.
The XIS was
operated in the normal clock/editing mode without an window
option. The target was observed at the XIS nominal
position. Table~\ref{table:xobslog} gives the log of the X-ray
observations.

\subsubsection{XIS data}

We analyze the XIS data in a standard manner with the
HEAsoft\footnote{http://heasrc.nasa.gov/lheasoft} version 6.6.2
software package. Photon events are extracted from a circular region
centered at the target with a radius of 6 mm ($4'.35$). The
background is taken from the surrounding annulus region. For spectral
analysis, we make the response matrix file (RMF) and ancillary
response file (ARF) using {\it xisrmfgen} and {\it xissimarfgen}
\citep{ishisaki2007}, respectively. The spectra of all the XIS sensors
are combined to improve the statistics, together with the
corresponding responses. We screen out events suffering from telemetry
saturation, using GTI filters provided by the Suzaku
team\footnote{http://www-cr.scphys.kyoto-u.ac.jp/menber/hiroya/xis/gtifile.html}.

\subsubsection{HXD/PIN data}

The HXD is a non-imaging collimated X-ray sensor consisting of two
main detection parts; silicon $p$-intrinsic-$n$ diodes (hereafter PIN)
and gadolinium silicate crystals (Ce-doped Gd$_2$SiO$_5$; GSO). They
cover the energy range of $10-70$~keV (PIN) and $40-600$~keV (GSO)
with low background levels \citep{takahashietal2007}. In the present
study, we concentrate on the PIN data, since the signals from SS~433
above 40 keV is below the sensitivity of the GSO. Since the
observations were performed in the normal mode of the HXD, we simply
extract the spectra from the PIN event data, following the standard
analysis
procedure\footnote{http://heasarc.gsfc.nasa.gov/docs/suzaku/analysis/abc/}.
We subtract the non X-ray background (NXB) from the data, utilizing
the modeled NXB distributed by the Suzaku team
\footnote{http://www.astro.isas.ac.jp/suzaku/analysis/hxd/pinnxb/}. The
version keyword of the NXB files is \verb|METHOD|=\verb|LCFITDT|
(i.e., so-called the ``tuned background''), which gives the best
reproductivity of the PIN NXB (0.34\% systematic error in the
15--40 keV for a 40 ksec exposure; \cite{fukazawaetal2009}).

\begin{figure}
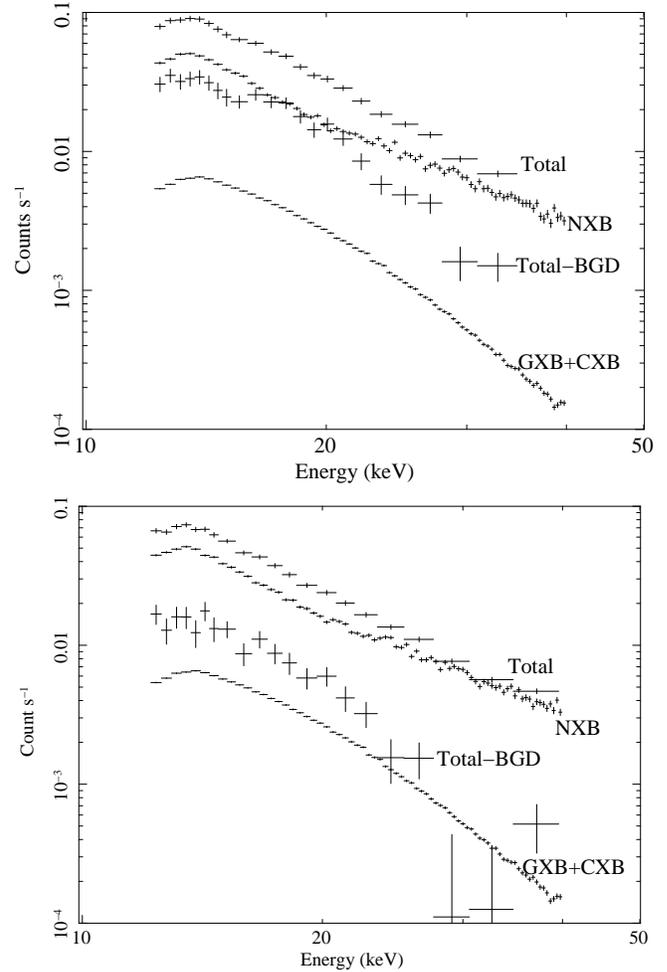

\begin{center}
 \includegraphics[angle=270,scale=0.35]{figure1a.ps}
 \includegraphics[angle=270,scale=0.35]{figure1b.ps}
\end{center}
\caption{
The PIN spectra of SS~433 on 2006 April 8--9 (upper, out-of-eclipse) and 
April 4--5 (lower, in-eclipse). 
From upper to lower, the total spectrum, the NXB, the 
source spectrum after background subtraction, and the GXB+CXB.
}
\label{fig:pinspec}
\end{figure}

\begin{figure}
\begin{center}
 \includegraphics[angle=270,scale=0.35]{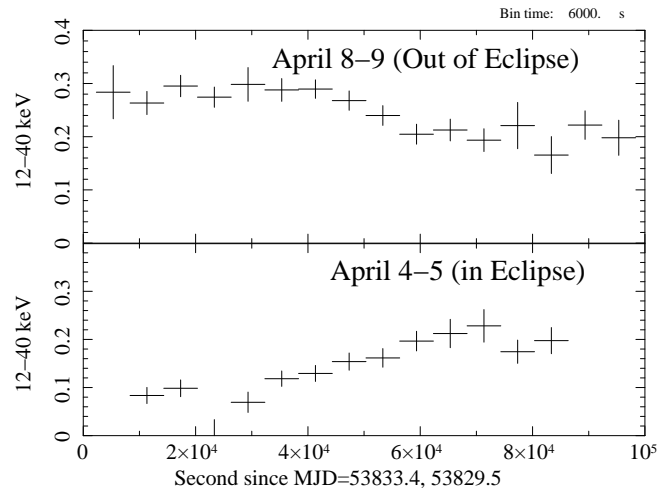}
\end{center}
\caption{
The PIN light curves of SS~433 in the 12--40 keV band on 2006 April
8--9 (upper, out-of-eclipse) and April 4--5 (lower, in-eclipse). The
NXB is subtracted.
}
\label{fig:lc_hxd}
\end{figure}

In addition to the NXB, two kinds of diffuse/extended background
should be subtracted before the spectral analysis; the cosmic X-ray
background (CXB) and the Galactic X-ray background (GXB). The GXB is
particularly important because a target source is close to the
Galactic plane.  Since the surface brightness of the GXB varies field
to field, we estimate its spectrum using the PIN data in a near-by
field, $(l,b)\sim(28^{\circ}.5,-0^{\circ}.2)$ (OBSID 500009020),
located $\approx$11$^{\circ}.5$ away from SS~433. This region is free
from bright X-ray sources and has been well studied with {\it
Chandra}, from which hundreds of faint point sources are detected
below 10~keV \citep{ebisawa2005,revnivtseveandsazonov2007}. They
cannot be resolved with the PIN, and hence contribute to the PIN
signal as a part of the GXB emission. To estimate the summed GXB+CXB
flux, we analyze the PIN spectra in this region by subtracting the NXB
in the same manner as above. We find the GXB+CXB spectrum can be well
fit with a single power-law model in the 12--40 keV band and obtain
the best-fit $\Gamma=2.1$ and flux of
$2.1\times10^{-11}$~erg~cm$^{-2}$~s$^{-1}$. Thus, assuming that the
GXB (and the CXB) is the same between this region and that around
SS~433, we subtract the modeled GXB+CXB spectrum from the PIN data of
SS~433, in addition to the NXB. Figure~\ref{fig:pinspec} shows the
total and source spectra of SS~433 after the background subtraction
for the eclipse and out-of-eclipse phase. The contribution of the NXB,
and the sum of the GXB and the CXB are separately plotted. We find
that the flux level of the GXB+CXB is only 0.3 times that of the NXB.
To evaluate the possible uncertainties in the estimated GXB flux,
we refer to the spatial distribution of the GXB obtained from the
Galactic ridge survey with the INTEGRAL/IBIS \citep{krivonos2007},
which covers almost the same energy band as the HXD/PIN. We confirm
that the systematic errors by using the different field separated by
10 degrees are smaller than the statistical errors, and hence do not
affect our conclusions. Figure~\ref{fig:lc_hxd} shows the light curves
of the HXD/PIN in the 12--40 keV from which the NXB is subtracted.

\subsection{Optical Spectroscopy}

We organized quasi-simultaneous optical spectroscopic observations of
SS~433 using ground-based telescopes around the epochs of the Suzaku
observations, aiming to trace the evolution of the jet structure from
the X-ray and optical data. From 2006 April 2 to 12, four
observatories participated in this campaign: the Special Astrophysical
Observatory (SAO), the Gunma Astronomical Observatory (GAO), the
Nishi-Harima Astronomical Observatory (NHAO), and the Asiago
Observatory of the University Padova. Table~\ref{table:optspeclog} gives the logs of
the optical observations. We observed SS~433 with the 6-m Russian
telescope\footnote{http://w0.sao.ru/Doc-en/Telescopes/bta/descrip.htm}
with the Multi Pupil Fiber Spectrograph
\footnote{http://w0.sao.ru/Doc-en/Telescopes/bta/instrum/}
at the SAO (PI: Fabrika). At the GAO, the 150 cm telescope equipped
with GLOWS\footnote{http://www.astron.pref.gunma.jp/instruments/device\_bcldsi.html} (Gunma LOW Resolution spectrograph and imager, $R$=400--500)
was used (PI: Kinugasa). At the NHAO, having the 2-m Nayuta
telescope\footnote{http://www.nhao.go.jp/nhao/telescopes/nayuta.html}, we
employed the optical spectrograph \citep{ozaki2005}
with a low dispersion mode ($R\sim$1000) for our observations (PI:
Ozaki). The Galileo 122-cm telescope was used at the Padova-Asiago
Observatory (PI: Iijima, $R$=800). We also took additional spectral data of
SS~433 at the GAO about 1-month before and 1 and 2-months after the
Suzaku observations. From these data, we determine the Doppler shift
of both receding and approaching jets using the H$\alpha$ emission
lines. Note that we were not able to detect H$\alpha$ originating from the
jets on April 2 (at GAO) and April 12 (at NHAO) due to the limited
signal-to-noise ratio.

\section{Suzaku Results I. (out of Eclipse)}\label{060408}

In this section, we present the results obtained from the Suzaku data
on 2006 April 8--9 (out of eclipse phase). For spectral fitting, we
use the XSPEC
package\footnote{http://heasarc.nasa.gov/docs/xanadu/xspec/} version
11.3.2ag \citep{xspec}.

\subsection{Light Curves}

\begin{figure}
\begin{center}
 \includegraphics[angle=270,scale=0.35]{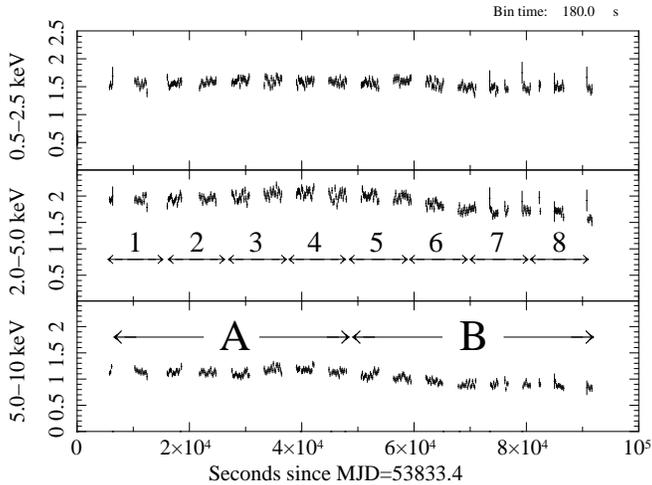}
\end{center}
\caption{
The XIS light curves of SS~433 on 2006 April 8--9 (out of eclipse) 
in three bands, 0.5--2 keV (upper), 2--5 keV (middle), and 
5--10 keV (lower). The bin size is 180 sec. The arrows define epochs~A and B
for the spectral analysis.
}
\label{fig:lc_out}
\end{figure}

\begin{figure}
\begin{center}
 \includegraphics[angle=270,scale=0.35]{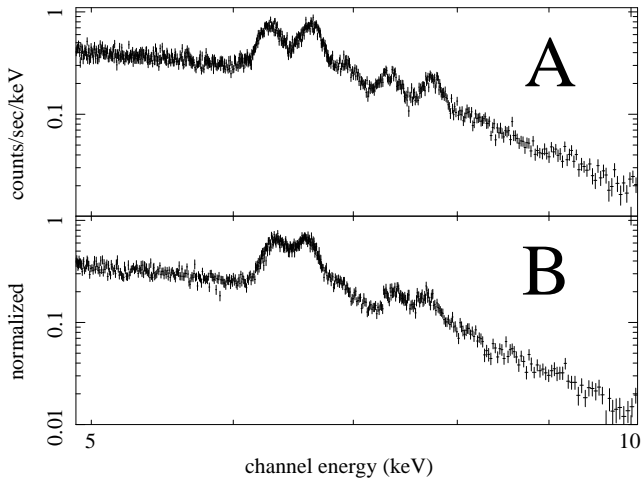}
\end{center}
\caption{The 5.0--10 keV spectra of SS~433 in epoch A (upper) and B
(bottom). The four XISs are summed. See Figure~\ref{fig:lc_out} for the
 definition of the epochs.}
\label{fig:hikaku}
\end{figure}

Figure~\ref{fig:lc_out} shows the summed XIS light curves of the four
sensors in the three energy bands, 0.5--2 keV, 2--5 keV, and 5--10 keV
for the XIS-FI (0.3--2 keV, 2--5 keV, and 5--8 keV for the XIS-BI).
As noticed, significant variability is found especially in the hard
band above the 5 keV.  We confirm that this tendency also appear in
hard X-ray band light curve in the 12--40 keV band obtained with the
HXD/PIN (see Figure~\ref{fig:lc_hxd}). For the subsequent
analysis, we divide the time region into two, epoch~A (the first
half), when the intensity and spectrum was almost stable, and epoch~B,
when the spectrum exhibited a drastic change. Figure~\ref{fig:hikaku}
shows the XIS spectra in epoch A and B. It is seen that the average
spectrum, in particular the iron-K emission line profiles, is very
different in the two epochs.

\subsection{Individual Spectral Analysis}

In this subsection, we study the properties of emission lines
originating from the blue and red jets and their evolution based on
only the XIS data. To examine the spectral variability, we divide the
one day data into eight parts (data number 1 to 8) with an equal
interval, and perform spectral fitting to each spectrum. To focus on
the iron-K band by making the best use of the energy resolution of the
XIS, we only analyze the 5.0--10 keV band.

The fitting model consists of a bremsstrahlung continuum and 11
emission lines, modified by the interstellar absorption. It is
expressed as a function of energy $E$:
\begin{displaymath}
\begin{array}{l} 
e ^{-\sigma (E) \, N_{\rm H}} \times [\mbox{Bremss}(kT)  + {\rm Fe\,
 _I\, K}\alpha_{z=0} \\
  + ( {\rm Fe\,_{XXV}\,K\alpha + Fe\,_{XXVI}\,K\alpha +
   Ni\,_{XXVII}\,K\alpha }\\
{\rm + Fe\,_{XXV} \,K\beta + Fe\, _{XXVI}\, K\beta} )_{z=z_{\rm red}}\\
  + ( {\rm Fe\, _{XXV} \,K\alpha + Fe\, _{XXVI} \,K\alpha + Ni\,
   _{XXVII} \,K\alpha }\\
{\rm + Fe\, _{XXV} \,K\beta + Fe\, _{XXVI} \,K\beta} )_{z=z_{\rm blue}}],
  \end{array}
\end{displaymath}
where $\sigma(E)$ is the photo-electric cross section and $N_{\rm H}$
is the hydrogen column density toward the source.  We fix $N_{\rm H}$
at $1.5\times10^{22}\,{\rm cm}^{-2}$, which is converted with the
formula by \citet{predehl1995} from $A_{V}$=8, a typical
extinction for SS~433 \citep{fabrika2004}.  In this study, for
simplicity, we adopt a single temperature bremsstrahlung model for the
continuum, considering the limited energy band; although a more
complicated model is applied for the analysis of the broad band
spectra in the next subsection, the results on the emission lines are
little affected.  The model includes 11 narrow Gaussian lines, five
pairs of Doppler-shifted lines from ${\rm Fe\,_{XXV}\,K\alpha}$, ${\rm
Fe\, _{XXVI}\, K\alpha}$, ${\rm Ni\,_{XXVII}\, K\alpha}$, ${\rm
Fe\,_{XXV}\, K\beta}$ and ${\rm Fe\,_{XXVI}\, K\beta}$ from the two
jets, and one stationary line of ${\rm Fe\, _I\, K}\alpha$.  The
Doppler shifts of the red and blue jets are free parameters that are
linked together among all the lines originating from the same jet.
The velocity dispersion (1$\sigma$ width) of the emission lines
is fixed at 1500 km s$^{-1}$, which is found by \citet{lopez2006} when
the precession phase is the closest to ours among available Chandra
HETGS observations.

First, we fit the first four spectra (epoch~A) simultaneously, which
exhibit little variability. In this fit, the intensities of the 6.4
keV ${\rm Fe\,_{I}\,K\alpha}$ lines are linked one another among all
the spectra. Then, we perform spectral fitting to the last four
spectra (epoch~B) separately. For these spectra, we fix the intensity
of the ${\rm Fe\,_{I}\,K\alpha}$ line at the best-fit value obtained
from the epoch~A data, $6.9\times10^{-5}\,{\rm
photons^{-1}\,cm^{-2}}$, since five emission lines (including four
jet lines) are blended around 6.4 keV and it is difficult to separate
its contribution in epoch~B.

\begin{figure}
\begin{center}
 \includegraphics[angle=270,scale=0.42]{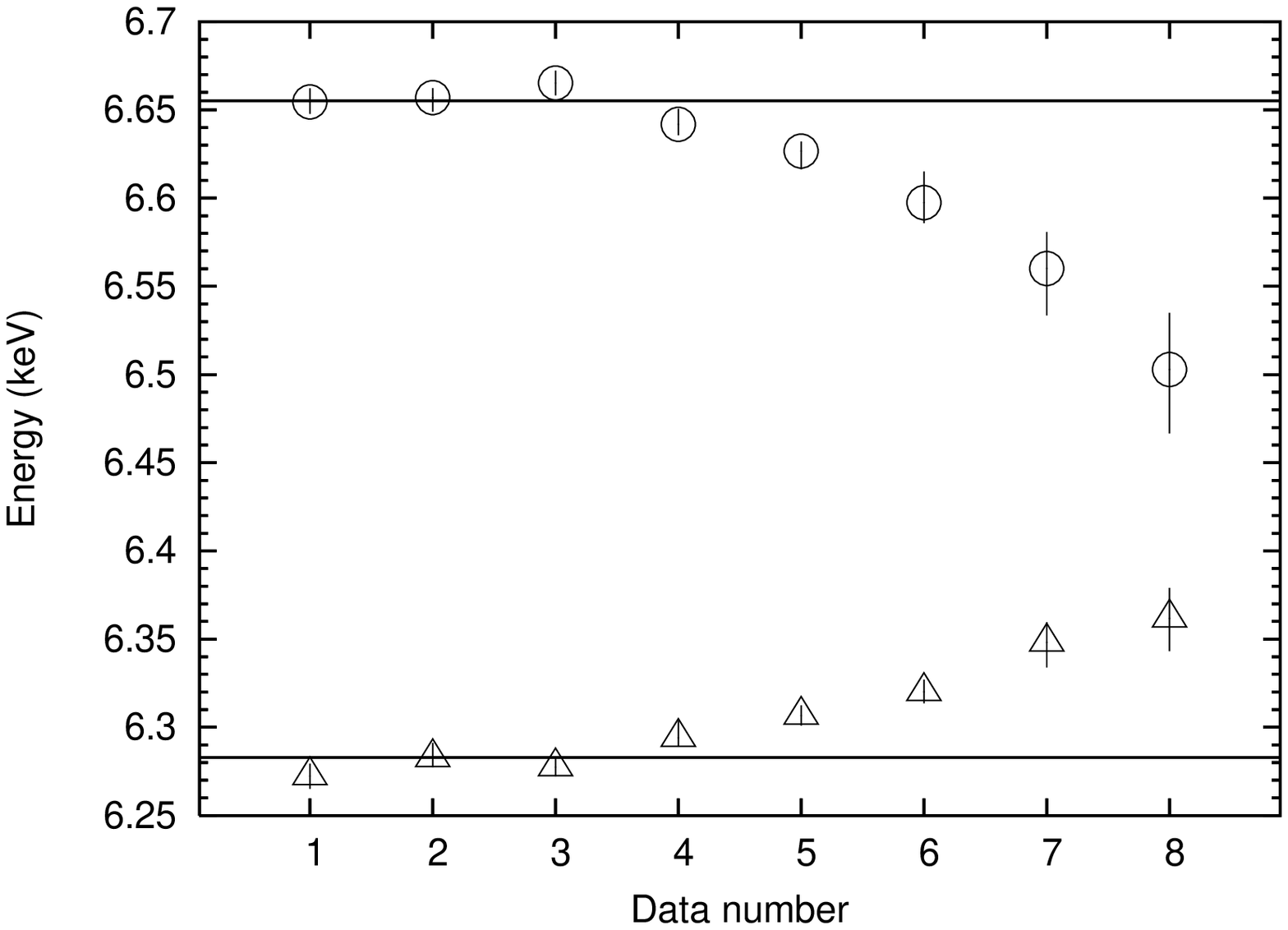}
\end{center}
\caption{Time variability of the line center energy of 
Fe$_{\rm XXV}\,$K$\alpha$ determined from the Suzaku spectra on 2008 April
8 (out-of-eclipse). The horizontal lines show the average values
determined from the time averaged spectra.}
\label{fig:z_out}
\end{figure}

Figure~\ref{fig:z_out} plots the time variability of the center
energies of the ${\rm Fe\,_{XXV}\,K\alpha}$ lines emitted from the
twin jets. As noticed, we detect rapid variability of the Doppler
shifts from data number 6 to 8, with a corresponding rate of ${\rm
d}z/{\rm d}t \approx$0.019/0.33 day$^{-1}$. Below we analyze the
summed spectra only from data number 1--4 (epoch~A), where the change
of the line energies is smaller than 20 eV.

\subsection{Broad Band Spectra}

We have obtained the simultaneous, broad band X-ray spectrum of SS~433
covering up to 40 keV with the XIS and HXD/PIN . This gives us an
ideal opportunity to investigate the origin of the continuum
emission. In this subsection, we perform a simultaneous fitting to the
XIS and PIN spectrum in the 5--10 keV and 12--40 keV band,
respectively. It is beyond the scope of this paper to apply a fully
analytic spectral model of the jets that includes both continuum and
line emission by self-consistent calculation. Instead, we only
consider a continuum model, over which the 11 emission lines are
superposed as independent parameters.

We adopt a multi-temperature bremsstrahlung model for the continuum;
assuming a power law dependence of the temperature of the jet $T$ on
the distance from the jet base $r$, we can compute a model spectrum
according to the expected relation between the differential emission
measure and temperature. Here we consider a stationary jet model
consisting of radiatively cooling, expanding plasmas (e.g.,
\cite{kotani1996}). The jets are assumed to move in a truncated cone
with a constant velocity. Hence, the density changes as $n(r)
\propto r^{-2}$. This model predicts approximately $T \propto
r^{-2(\gamma-1)}$, where $\gamma$ is the adiabatic index (5/3 for non
relativistic particles). Since we cannot well constrain the power law
index from our data, we assume $T\propto r^{-4/3}$ according to this
model. The free parameters are the innermost temperature $T_0$ and its
normalization.

We also consider a reflection component from cold matter, which is
indicated by the presence of the fluorescence 6.4 keV iron K emission
line, although it has been neglected in most of previous studies. For
this we adopt the reflection code by \citet{magdziarz1995}, available
as {\it reflect} model on XSPEC applicable for any incident
spectrum. The free parameter is the the reflection strength $R
\equiv\Omega/2\pi$, where $\Omega$ is the solid angle seen by the
X-ray emitter (i.e., the jets). We fix solar metal abundances and an
inclination angle of 79$^\circ$, assuming that the reflection mainly
takes place in outer parts of the accretion disk.

\begin{figure}
\begin{center}
 \includegraphics[angle=270,scale=0.35]{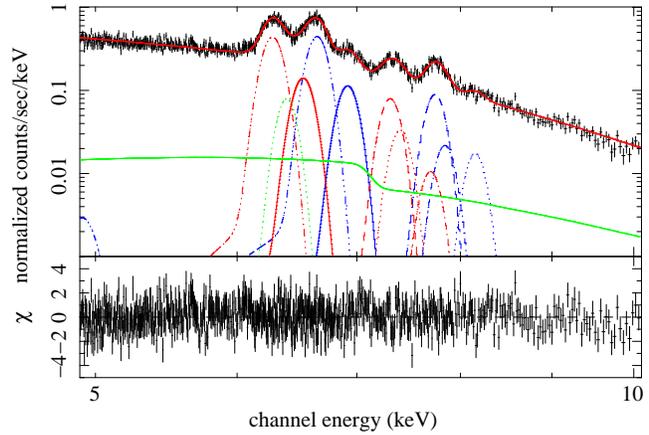}
\end{center}
\caption{The XIS spectrum in the 5--10 keV band on 2006 
April 8--9 (data number 1--4 only; out of eclipse). 
The best-fit model is over-plotted with separate contribution 
of the reflection component (lower curve, green) and Gaussians 
(blue: those from the blue jet, red: those from the red jet, 
green: the fluorescence iron-K line at 6.4 keV).
The fitting residuals in units of $\chi$ are plotted in the lower
panel.  The best fit parameters are given in Table~\ref{table:data}.}
\label{fig:spec_out}
\end{figure}

\begin{figure}
\begin{center}
 \includegraphics[angle=270,scale=0.35]{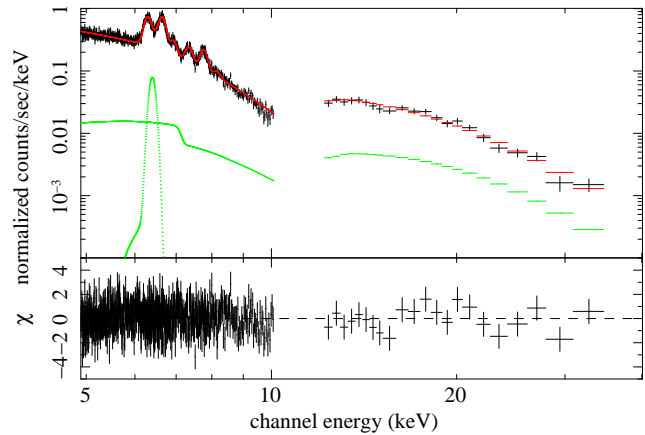}
\end{center}
\caption{
The simultaneous fit to the XIS and HXD/PIN spectra of 2006 April 8--9
(out-of-eclipse), using a multi-temperature jet
model for the continuum (see text). 
The best-fit models are over-plotted with 
separate contribution from the reflection component 
and fluorescence iron-K line (green).
Residuals in units of $\chi$ are shown in the lower panel. See
Table~\ref{table:data} for the best-fit parameters of the continuum.}
\label{fig:hxd_out}
\end{figure}

The fitting results for the epoch~A spectra are shown in
Figures~\ref{fig:spec_out} and \ref{fig:hxd_out}. The best-fit
parameters are summarized in Table~\ref{table:data}. In the analysis,
the relative normalization between the XIS and HXD/PIN is fixed at
0.923, based on the calibration using the Crab Nebula
\citep{ishida2007}. We obtain the innermost jet temperature of $T_0
=27\pm2$ keV. To make the reflection parameter physically meaningful,
we limit $R \leq 1$ in the fitting and finally obtain
$R=1.0^{+0}_{-0.41}$. The equivalent width of the 6.4 keV line with
respect to the reflection component turns out to be 1.0 keV at
the best-fit value, well consistent with theoretical calculations
\citep{Matt1991} within a possible range of viewing angle and an iron
abundance.

\section{Suzaku Results II. (in Eclipse)}\label{060404}

\begin{figure}
\begin{center}
 \includegraphics[angle=270,scale=0.35]{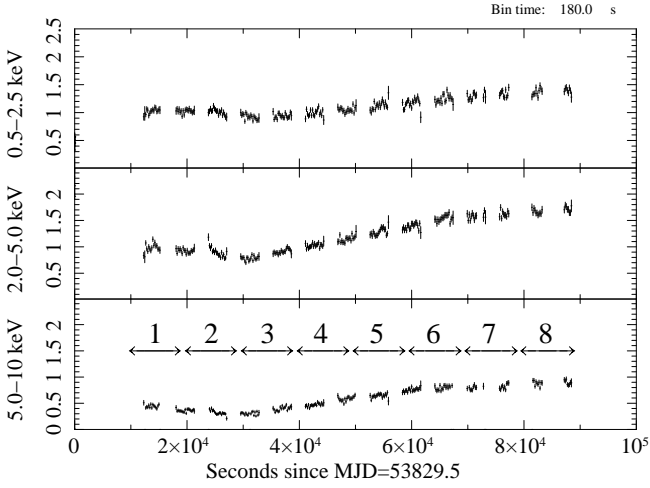}
\end{center}
\caption{
The XIS light curves of SS~433 on 2006 April 4--5 (in eclipse) 
in three bands, 0.5--2 keV (upper), 2--5 keV (middle), and 
5--10 keV (lower). The bin size is 180 sec.}
\label{fig:lc_in}
\end{figure}

\begin{figure}
\begin{center}
 \includegraphics[angle=270,scale=0.42]{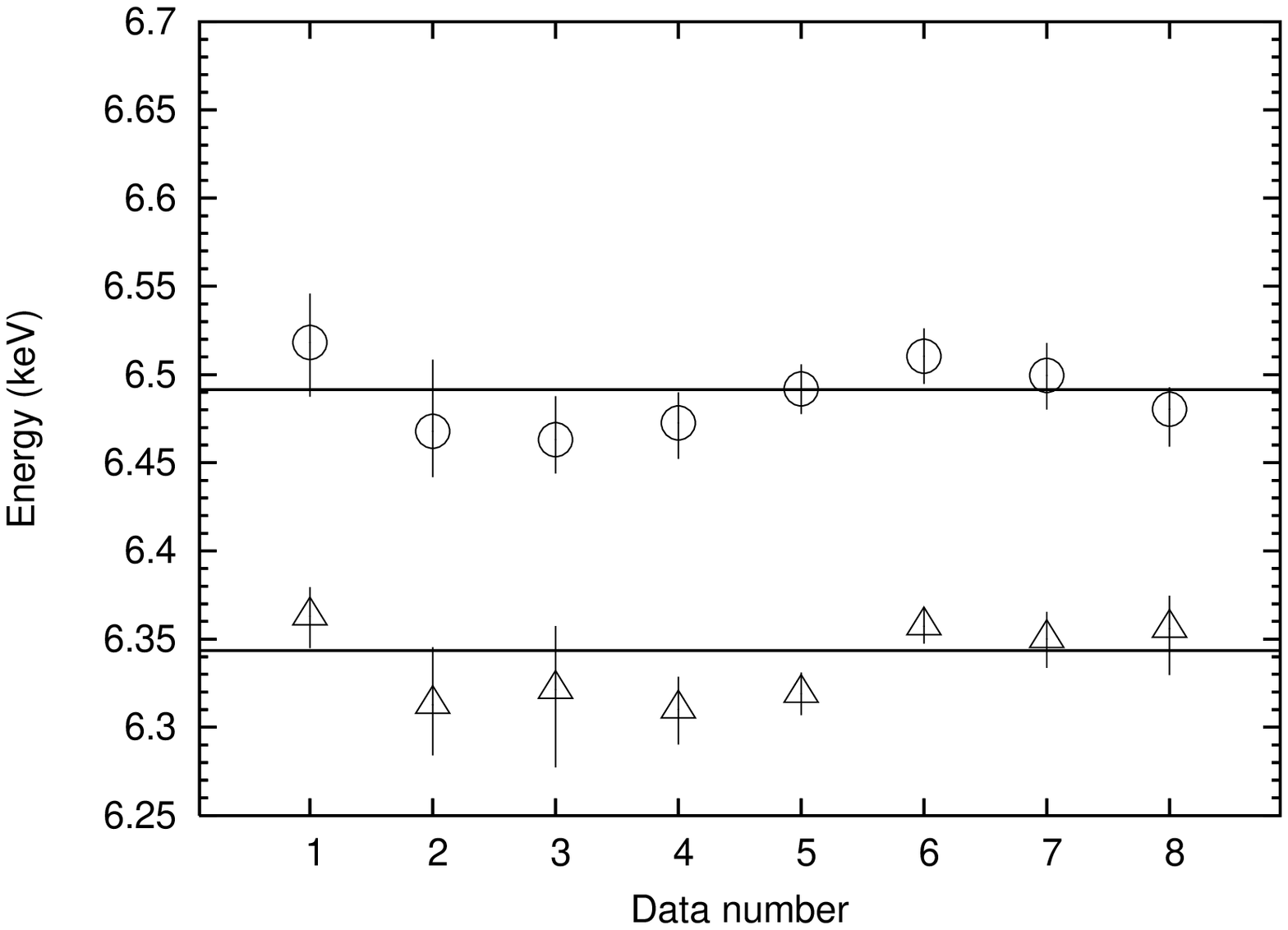}
\end{center}
\caption{
Time variability of the line center energy of Fe$_{\rm
XXV}\,$K$\alpha$ determined from the Suzaku spectra on 2008 April
4--5 (in eclipse). The horizontal lines show the average values
determined from the time averaged spectra.
}
\label{fig:z_in}
\end{figure}

The same analysis as in the previous section is applied for the data
of the first Suzaku observation (in-eclipse). Figure~\ref{fig:lc_in}
shows the X-ray light curves on 2006 April 4--5 in the three energy
bands. A long term variability is clearly seen, as expected from the
eclipse of the jets by the companion star. Similarly to the
out-of-eclipse data, we make 8 time-sliced spectra and perform
spectral fitting to each one, using the same spectral model described
in \S~3. In this period, since the spectra around 6.4 keV are very
crowded and hence the intensity of the Fe$_{\rm I}$ K$\alpha$ line
cannot be determined, we fix it at $3.7\times 10^{-5}\,{\rm
photons^{-1}\,cm^{-2}}$, a value that is estimated from the analysis
of the broad-band spectra with a reflection component, as described
below. Figure~\ref{fig:z_in} shows the evolution of the line center
energies of the ${\rm Fe\,_{XXV}\,K\alpha}$ lines. We do not detect
significant variability of the Doppler shifts which were observed in
the out-of-eclipse phase (Figure~\ref{fig:z_out}).

\begin{figure}
\begin{center}
 \includegraphics[angle=270,scale=0.35]{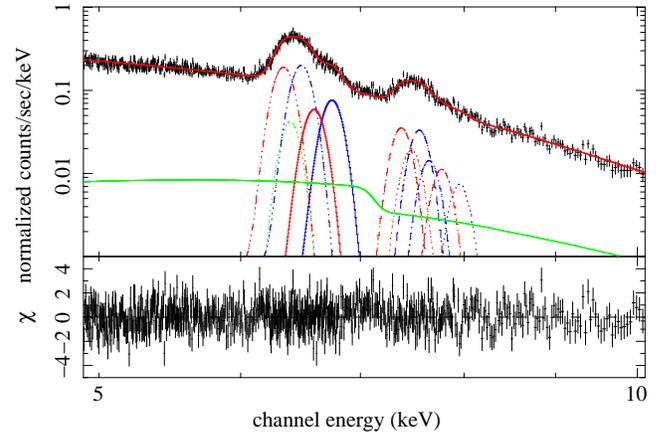}
\end{center}
\caption{
The XIS spectrum in the 5--10 keV band on 2006 April 4--5 (in eclipse). 
The best-fit model is over-plotted with separate contribution 
of the reflection component (lower curve, green) and Gaussians 
(blue: those from the blue jet, red: those from the red jet, 
green: the fluorescence iron-K line at 6.4 keV).
The fitting residuals 
in units of $\chi$ are plotted in the lower panel. 
The best fit parameters are given in Table~\ref{table:data}.
}
\label{fig:spec_in}
\end{figure}

\begin{figure}
\begin{center}
 \includegraphics[angle=270,scale=0.35]{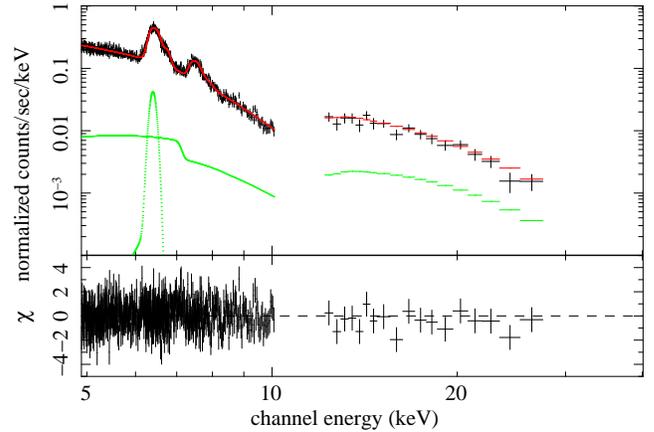}
\end{center}
\caption{
The simultaneous fit to the XIS and HXD/PIN spectra of 2006 April 4--5
(in eclipse), using a multi-temperature jet
model for the continuum (see text). 
The best-fit models are over-plotted with 
separate contribution from the reflection component 
and fluorescence iron-K line (green).
Residuals in units of $\chi$ are
shown in the lower panel. See Table~\ref{table:data} for the best-fit
 parameters of the continuum.}
\label{fig:hxd_in}
\end{figure}

We also make simultaneous fit to the XIS and HXD/PIN spectra to
constrain the continuum model, using the same multi-temperature jet
model with a reflection component as described in the previous
subsection.  We limit $R\leq 1$ and fix the equivalent width of the
fluorescence Fe$_{\rm I}$ K$\alpha$ line with respect to the
reflection component at 1.0 keV, based on the result of the
out-of-eclipse data, since it is difficult to constrain the absolute
intensity from the data. We obtain $T_0 = 21 \pm 1$ keV and
$R=1.0^{+0}_{-0.13}$.  Figures~\ref{fig:spec_in} and \ref{fig:hxd_in}
show the data and best-fit model. The best fit parameters are
summarized in Table~\ref{table:data}.

\section{Discussion}

\subsection{X-ray and Optical Doppler Shift Curves}

\begin{figure}
\begin{center}
 \includegraphics[angle=270,scale=0.42]{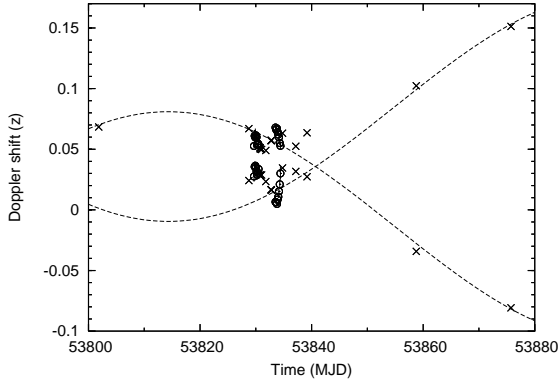}
\end{center}
\caption{A summary of Doppler shifts of the jets observed around
the observation campaign epoch. The results
 determined from the optical (H$\alpha$) and X-ray (Fe$_{\rm XXV}\,$K$\alpha$)
are shown by the crosses and circles, respectively.
The doted lines correspond to the 162.15-days sinusoidal curves due to the 
precession motions.} 
\label{fig:doppler}
\end{figure}

Figure~\ref{fig:doppler} shows the Doppler shifts of the SS~433 jets
measured with the Suzaku XIS (circles) and optical spectroscopy
(crosses). The two dot lines correspond to the 162.15 days period
sinusoidal curves for the blue and red jets expected from the {\it
precession} motion \citep{gies2002}, which is shifted by
--5.5 days to fit the observational points. From this figure, it is
confirmed that the Suzaku observations were performed at a precession
phase when the jets axis are almost perpendicular to the line of
sight.

\begin{figure}
\begin{center}
 \includegraphics[angle=270,scale=0.42]{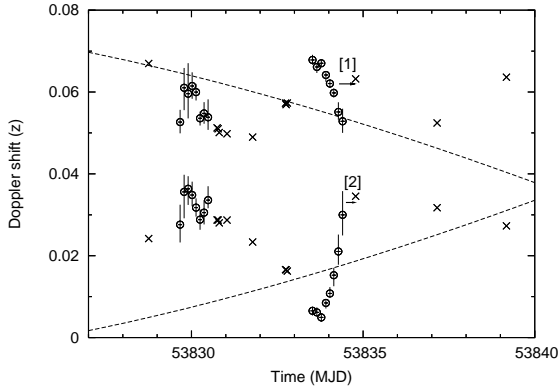}
\end{center}
\caption{A blow up of Figure~\ref{fig:doppler} around the epoch 
of the Suzaku observations.}
\label{fig:doppler_can}
\end{figure}

Figure~\ref{fig:doppler_can} is a blow-up of Figure~\ref{fig:doppler}
around the Suzaku observation epoch. It is clearly seen that the
observed Doppler shift curves both from the X-ray and optical data
show strong deviation from the precession curve with a 162.15 days
period. Except for the data around 2006 April 8 (MJD 53834), both
amplitude and periodicity of this deviation are consistent with the
{\it nodding} motion of the jets, expressed as a sinusoidal curve with
a 6.28 days period and a semi-amplitude of $\Delta z
\approx$ 0.012 at our precession phase \citep{fabrika2004}.

As mentioned above, in the Suzaku XIS data on 2006 April 8, we detect
rapid variability of the Doppler shifts, ${\rm d}z/{\rm d}t\approx$
0.019/0.33 day$^{-1}$. This is larger than those expected from the
precession and/or nodding motion of the jets; the maximum variability
from the nodding motion is ${\rm d}z/{\rm d}t \approx$0.004/0.33
day$^{-1}$ \citep{fabrika2004}. Hence, we interpret that our Suzaku
result corresponds to a {\it jitter} which was previously reported
only from the optical spectroscopy (e.g., \cite{iijima1993}), observed
for the first time in X-rays.  The origin of these phenomena will be
discussed in the next subsection.

Utilizing our quasi simultaneous X-ray and optical spectroscopic data,
we can determine a time lag between the epochs when same jet material
emits X-rays and optical lights, assuming that the Doppler shift
remains constant in traveling along the jet. Based on the optical data
points on MJD 53834.8 and previous X-ray curves, the lag in the
red and blue jet is estimated to be 0.64$\pm$0.08 day and
0.38$\pm$0.08 day, respectively. Assuming that the jets speed is
constant at $0.26 c$, we estimate the distance of the optical jets
measured from the X-ray jets to be $l_{\rm opt} = (4.3\pm0.6)\times
10^{14}$ cm (red) and $l_{\rm opt} = (2.6\pm0.6)\times 10^{14}$
cm (blue). Thus, the H$\alpha$ emitting region may not be
at the symmetric location between the twin jets, depending on the
efficiency of the cooling determined by the plasma density. The order
of the jet size is well consistent with previous estimates
\citep{fabrika2004}.

\begin{figure}
\begin{center}
 \includegraphics[angle=270,scale=0.42]{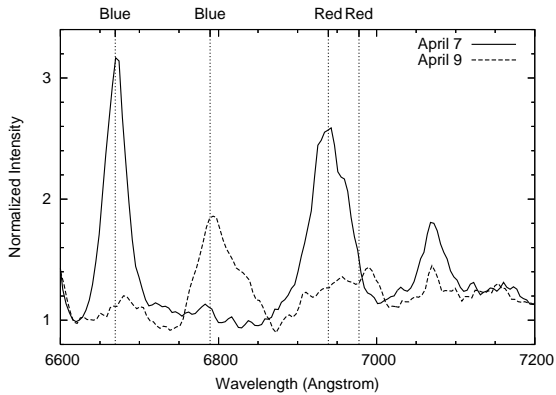}
\end{center}
\caption{The optical spectra of SS~433 observed at the GAO 
on 2006 April 7 (solid line) and April 9 (dashed line).
The vertical lines denote the line center positions of H$\alpha$ 
originating from the red and blue jets. There are two
He$_{\rm I}$ lines ($\lambda$6678$, \lambda$7065) in the figure. 
}
\label{fig:opt_spec}
\end{figure}

The jitter observed in X-rays on 2006 April 8--9 is confirmed in the
optical spectra taken after this event. Since the length of the
optical jets is 1000 times larger than that of the X-ray jets, fast
time variability observed in X-rays is significantly smeared out in
the optical band. Thus, soon after a rapid change of Doppler shifts is
observed in X-rays, the optical emission lines become broad (if
unresolved), by including the contribution from jet material ejected
at different epochs.
This is called a ``projection effect'', previously observed from nodding
motions \citep{borisov1987}. 
 Thus, we investigate the change of the width of
H$\alpha$ lines originating from the jets before and after the second
Suzaku observation (out-of-eclipse). The optical spectra observed at
the GAO are plotted in Figure~\ref{fig:opt_spec}. We find that while
the H$\alpha$ line width of the blue and red jets was 40 
\AA\ (FWHM)
and 58 \AA\ on April 7, the lines became significantly broader with a
FWHM of 188 \AA\ and 315 \AA, respectively, on April 9. This change of
the H$\alpha$ line width corresponds to $\Delta z$=0.023 (blue jet) and
$\Delta z$=0.039 (red jet), which is larger only by $\Delta z$=0.005
than the total observed change of the Doppler shifts of X-ray
lines. Thus, the broadening of the H$\alpha$ lines seen in the April 9
spectrum is mainly attributable to the rapid change of the Doppler
shifts occurring close to the jet base.

\subsection{Origin of the X-ray {\it Jitter}}

\begin{figure}
\begin{center}
 \includegraphics[angle=270,scale=0.42]{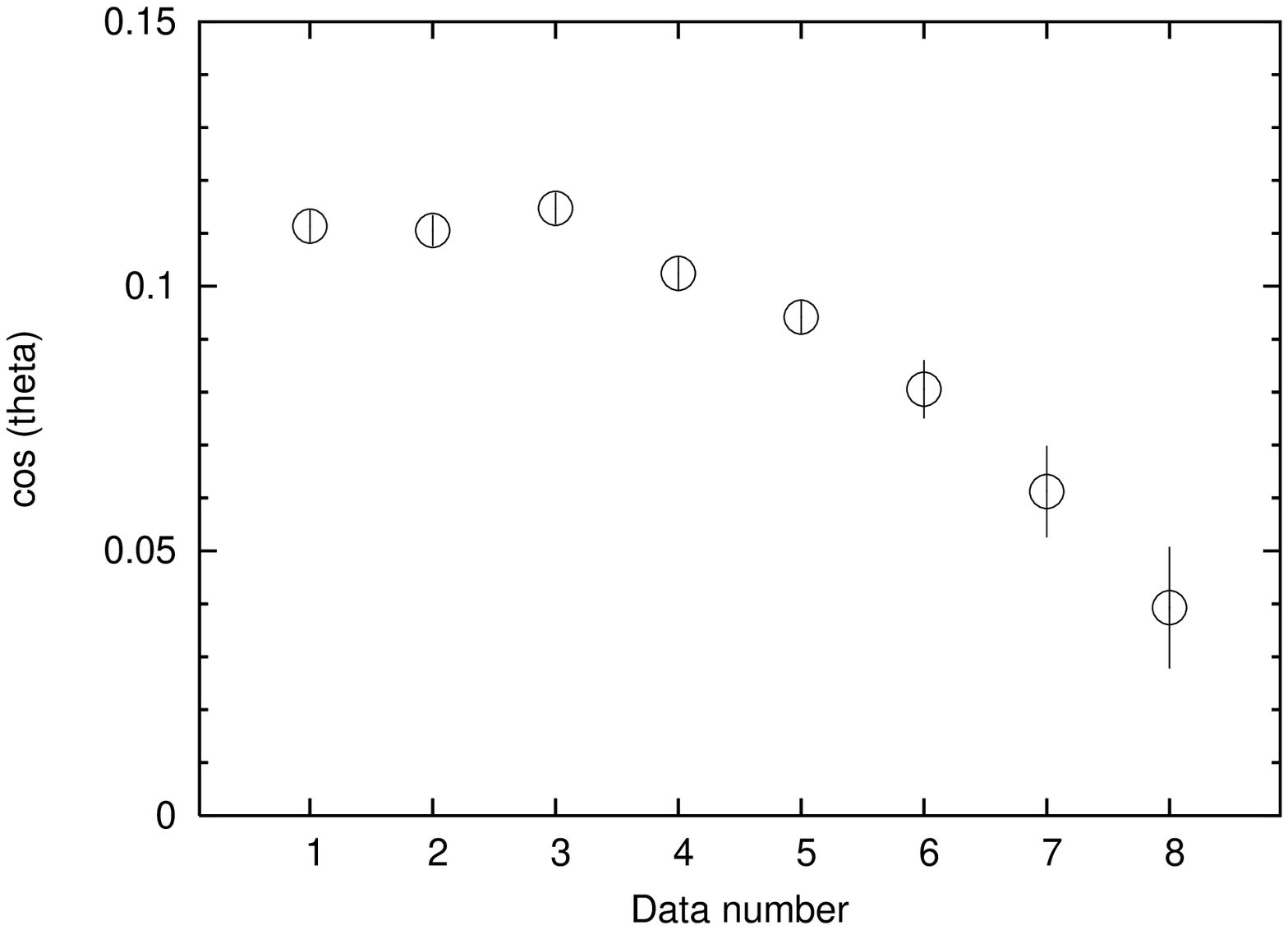}
 \includegraphics[angle=270,scale=0.42]{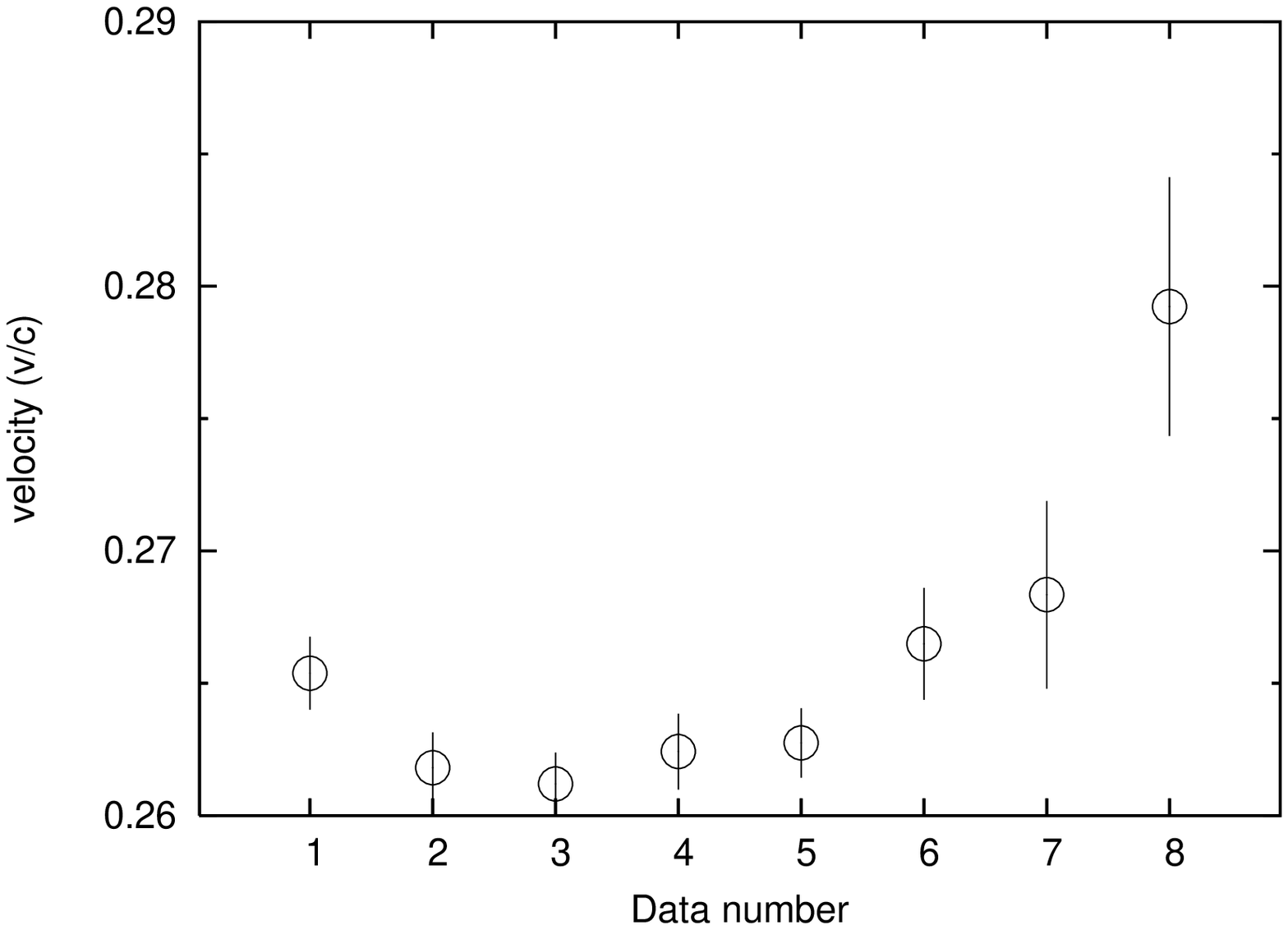}
\end{center}
\caption{The change of the jet parameters observed on 2006 April 8--9, 
in terms of the cosine of angle between the line-of-sight and jet axis (upper)
and intrinsic velocity (lower).
}
\label{fig:jet_parameters}
\end{figure}

We have detected the fastest variability of Doppler shifts of X-ray
emission lines ever observed from SS~433, which can be interpreted as
{\it jitter} motions of the jets, not the nodding motions where
much slower change of redshifts is expected. Assuming that the twin
jets are always symmetric, we can estimate the orientation and
intrinsic speed of the jets from the observed Doppler shifts of blue
and red components. The results are plotted in
Figure~\ref{fig:jet_parameters}. As noticed from the figure, we detect
a significant change in both angle and speed of the jets during the
latter interval. From the former half toward the end of the
observation, the jet speed increased by a factor of
$(8\pm2)$\%. Recall that in the same epoch, the X-ray spectra became
softer and hence the temperature of the jet decreased. This
anti-correlation between the jet speed and temperature must be
explained in theoretical models of jets.

Considering the smooth evolution of flux and line center energy,
we infer that the X-ray jitter originates from a {\it continuous}
change of the jet kinetic parameters, not reflecting the production of
discrete jets with largely different Doppler shifts. Since the length
of the X-ray jets is $\sim 10^{12}$ cm corresponding to a traveling
time of only $\sim 100$ sec, we simply measure the flux-weighted
average of Doppler shifts of jets emitted during the integration time
($\sim 10^4$ sec) of each spectrum. If jets with very different
Doppler shifts are emitted within each exposure, we would expect
broadening of the emission lines. We check the line widths of the
$\rm Fe\,_{XXV}\,K\alpha$ emission lines from the 8 time-sliced
spectra on 2006 April 8--9. The same fitting model as in
\S~\ref{060408} is used, by making the width of Gaussian lines
free. We find that the width is almost constant over the observation
at $34\pm6$ eV (the average from the blue jet) and $36\pm7$ eV (red
jet), which are comparable to the observed line width resolved with
Chandra/HETGS \citep{namiki2003,lopez2006}. No significant line
broadening is observed in epoch~B. Thus, we favor the continuously
changing jet model, although we cannot exclude the possibility of many
discrete jets with slightly different kinetic parameters. In the
optical jets, by contrast, the evolution of emission lines from
discrete plasmoids were sometimes observed on a time scale of
$\sim$0.1--0.3 days \citep{panferov1993}. The relation of these
phenomena to the X-ray jitter is not clear.

\subsection{Broad Band X-ray Spectra}

We obtain for the first time simultaneous broad band spectra of SS~433
covering up to $\sim$40 keV including those observed with CCD
resolution below 10 keV with Suzaku. This gives us a unique
opportunity to investigate the origin of the spectra not only from the
emission lines but also from the continuum shape. We find that the
spectrum in eclipse is consistent with the emission from optically
thin plasmas in the twin jets as reported in previous works
\citep{brinkmann1991,kotani1996}, although the out-of-eclipse spectrum
may need an additional hard component to self-consistently explain
both emission lines and continuum.

From the line intensities, we can estimate the innermost temperature
of the jets based on the radiatively cooling jet model, as done in
\citet{kotani1996}. The innermost temperature derived from the ratio
of the Fe$_{\rm XXV}$ and Fe$_{\rm XXVI}$ K$\alpha$ emission lines are
summarized in Table~\ref{table:data} as ``Jet Temperature'' $T_0^{\rm
line}$. The averaged temperature of the blue and red jets is found to
be $\approx$13 keV and $\approx$16 keV, respectively. Due to the
confusion with the 6.4 keV line under the limited energy resolution,
these could be a systematic uncertainty in the line intensity of
Fe$_{\rm XXV}$. We conservatively estimate the possible systematic
errors in the line ratio of Fe$_{\rm XXV}$ and Fe$_{\rm XXVI}$ to be
$\approx$10\%, which corresponds to $\approx$15\% (2--3 keV) in the
innermost temperature. If we take into account this uncertainty, the
averaged innermost temperature is consistent with each other between
the out-of-eclipse and in-eclipse phase. In our precession phase
(edge-on), the region size obscured by the outer rim of the accretion
disk and that by the companion star is similar, and hence there is no
large difference in $T_0^{\rm line}$ between the two Suzaku observations.

Independently, we are able to constrain the innermost temperature
$T_0$, listed in Table~\ref{table:data}, directly from the continuum
shape based on the same jet model, thanks to the high quality hard
X-ray spectra obtained with the HXD/PIN. We find that in the eclipse
phase, this $T_0$ value (21$\pm$1 keV) is close to
$T_0^{\rm line}$ (16$\pm$3 keV, the average of the blue and red jets).
In the out-of-eclipse phase, however, $T_0$
determined from the continuum (27$\pm$2 keV) is significantly higher
than that determined from the Fe line ratio (13$\pm$2
keV). These facts suggest that the out-of-eclipse spectrum may be
contaminated by an additional hard X-ray component that is hidden in
eclipse. The spectrum must be very hard and/or heavily absorbed since
its contribution should not be large below $\sim$7 keV, where the
ratio of the continuum intensity between the out-of-eclipse and
in-eclipse phases is similar to that of the iron emission lines
originating from the jets. The presence of such a hard component not
directly related to the jets was suggested by the comparison of in-
and out-of-eclipse spectra observed with Ginga
\citep{brinkmann1991,yuan1995}.  This may be related to the extended
corona around the accretion disk discussed in \citet{krivosheyev2009}
or a reflection feature of the emission from the inner disk by the
outer wall of the funnel \citep{medvedev2009}.  To constrain its
origin, analysis of broad band data utilizing a fully self-consistent
model for the thermal jets (including both continuum and emission
lines) would be required, which we leave for future studies.

\section*{Acknowledgment}

This work was partly supported by the Grant-in-Aid for JSPS Fellows
for young researchers (KK), the Grants-in-Aid for Scientific Research
20540230 (YU), and the Grant-in-Aid for the Global COE Program ``The
Next Generation of Physics, Spun from Universality and Emergence''
from from the Ministry of Education, Culture, Sports, Science and
Technology (MEXT) of Japan.

\bibliographystyle{apj}
\bibliography{biblist}

\begin{thebibliography}{44}
\expandafter\ifx\csname natexlab\endcsname\relax\def\natexlab#1{#1}\fi

\bibitem[{{Arnaud} {et~al.}(2006){Arnaud}, {Dorman}, \& {Gordon}}]{xspec}
{Arnaud}, K., {Dorman}, B., \& {Gordon}, C. 2006, XSPEC: An X-ray Spectral
  Fitting Package User's Guide for 12.3.0, HEASARC

\bibitem[{{Borisov} \& {Fabrika}(1987)}]{borisov1987}
{Borisov}, N.~V., \& {Fabrika}, S.~N. 1987, Soviet Astronomy Letters, 13, 200

\bibitem[{{Brinkmann} {et~al.}(1991){Brinkmann}, {Kawai}, {Matsuoka}, \&
  {Fink}}]{brinkmann1991}
{Brinkmann}, W., {Kawai}, N., {Matsuoka}, M., \& {Fink}, H.~H. 1991, \aap, 241,
  112

\bibitem[{{Brinkmann} {et~al.}(2005){Brinkmann}, {Kotani}, \&
  {Kawai}}]{brinkmann2005}
{Brinkmann}, W., {Kotani}, T., \& {Kawai}, N. 2005, \aap, 431, 575

\bibitem[{{Cherepashchuk} {et~al.}(2007){Cherepashchuk}, {Sunyaev}, {Seifina},
  {Antokhina}, {Kosenko}, {Molkov}, {Shakura}, {Postnov}, {Timokhin}, \&
  {Panchenko}}]{cherepashchuk2007}
{Cherepashchuk}, A.~M. et al. 2007, in ESA Special Publication,
  Vol. 622, ESA Special Publication, 319

\bibitem[{{Cherepashchuk} {et~al.}(2005){Cherepashchuk}, {Sunyaev}, {Fabrika},
  {Postnov}, {Molkov}, {Barsukova}, {Antokhina}, {Irsmambetova}, {Panchenko},
  {Seifina}, {Shakura}, {Timokhin}, {Bikmaev}, {Sakhibullin}, {Aslan},
  {Khamitov}, {Pramsky}, {Sholukhova}, {Gnedin}, {Arkharov}, \&
  {Larionov}}]{cherepashchuk2005}
{Cherepashchuk}, A.~M. st al. 2005, \aap, 437, 561

\bibitem[{{Ciatti} {et~al.}(1981){Ciatti}, {Mammano}, \&
  {Vittone}}]{ciatti1981}
{Ciatti}, F., {Mammano}, A., \& {Vittone}, A. 1981, \aap, 94, 251

\bibitem[{{Collins} \& {Garasi}(1994)}]{collins1994}
{Collins}, II, G.~W., \& {Garasi}, C.~J. 1994, ApJ, 431, 836

\bibitem[{{Ebisawa} {et~al.}(2005){Ebisawa}, {Tsujimoto}, {Paizis},
  {Hamaguchi}, {Bamba}, {Cutri}, {Kaneda}, {Maeda}, {Sato}, {Senda}, {Ueno},
  {Yamauchi}, {Beckmann}, {Courvoisier}, {Dubath}, \&
  {Nishihara}}]{ebisawa2005}
{Ebisawa}, K. et al. 2005, \apj, 635, 214

\bibitem[{{Fabrika}(2004)}]{fabrika2004}
{Fabrika}, S. 2004, Astrophysics and Space Physics Reviews, 12, 1

\bibitem[{{Fukazawa} {et~al.}(2009){Fukazawa}, {Mizuno}, {Watanabe}, {Kokubun},
  {Takahashi}, {Kawano}, {Nishino}, {Sasada}, {Shirai}, {Takahashi}, {Umeki},
  {Yamasaki}, {Yasuda}, {Bamba}, {Ohno}, {Takahashi}, {Ushio}, {Enoto},
  {Kitaguchi}, {Makishima}, {Nakazawa}, {Uehara}, {Yamada}, {Yuasa}, {Isobe},
  {Kawaharada}, {Tanaka}, {Tashiro}, {Terada}, \& {Yamaoka}}]{fukazawaetal2009}
{Fukazawa}, Y. et al. 2009, \pasj, 61, 17

\bibitem[{{Gies} {et~al.}(2002){Gies}, {McSwain}, {Riddle}, {Wang}, {Wiita}, \&
  {Wingert}}]{gies2002}
{Gies}, D.~R., {McSwain}, M.~V., {Riddle}, R.~L., {Wang}, Z., {Wiita}, P.~J.,
  \& {Wingert}, D.~W. 2002, ApJ, 566, 1069

\bibitem[{{Goranskii} {et~al.}(1998){Goranskii}, {Esipov}, \&
  {Cherepashchuk}}]{goranskii1998}
{Goranskii}, V.~P., {Esipov}, V.~F., \& {Cherepashchuk}, A.~M. 1998, Astronomy
  Reports, 42, 209

\bibitem[{{Iijima}(1993)}]{iijima1993}
{Iijima}, T. 1993, ApJ, 410, 295

\bibitem[{{Ishida} {et~al.}(2007){Ishida}, {Suzuki}, \& {Someya}}]{ishida2007}
{Ishida}, M., {Suzuki}, K., \& {Someya}, K. 2007, JX-ISAS-SUZAKU-MEMO-2007-11

\bibitem[{{Ishisaki} {et~al.}(2007){Ishisaki}, {Maeda}, {Fujimoto}, {Ozaki},
  {Ebisawa}, {Takahashi}, {Ueda}, {Ogasaka}, {Ptak}, {Mukai}, {Hamaguchi},
  {Hirayama}, {Kotani}, {Kubo}, {Shibata}, {Ebara}, {Furuzawa}, {Iizuka},
  {Inoue}, {Mori}, {Okada}, {Yokoyama}, {Matsumoto}, {Nakajima}, {Yamaguchi},
  {Anabuki}, {Tawa}, {Nagai}, {Katsuda}, {Hayashida}, {Bamba}, {Miller},
  {Sato}, \& {Yamasaki}}]{ishisaki2007}
{Ishisaki}, Y. et al. 2007, \pasj,
  59, 113

\bibitem[{{Katz} \& {Piran}(1982)}]{katz1982}
{Katz}, J.~I., \& {Piran}, T. 1982, \aplett, 23, 11

\bibitem[{{Kawai} {et~al.}(1989){Kawai}, {Matsuoka}, {Pan}, \&
  {Stewart}}]{kawai1989}
{Kawai}, N., {Matsuoka}, M., {Pan}, H.-C., \& {Stewart}, G.~C. 1989, \pasj, 41,
  491

\bibitem[{{Kotani} {et~al.}(2008){Kotani}, {Fabrika}, {Goranskij}, {Kawai},
  {Kinugasa}, {Kubota}, {Nakanishi}, {Trushkin}, \& {Tsuboi}}]{kotani2008}
{Kotani}, T. et al. 2008, in VII
  Microquasar Workshop: Microquasars and Beyond

\bibitem[{{Kotani} {et~al.}(1994){Kotani}, {Kawai}, {Aoki}, {Doty}, {Matsuoka},
  {Mitsuda}, {Nagase}, {Ricker}, \& {White}}]{kotani1994}
{Kotani}, T. et al. 1994, \pasj, 46, L147

\bibitem[{{Kotani} {et~al.}(1996){Kotani}, {Kawai}, {Matsuoka}, \&
  {Brinkmann}}]{kotani1996}
{Kotani}, T., {Kawai}, N., {Matsuoka}, M., \& {Brinkmann}, W. 1996, \pasj, 48,
  619

\bibitem[{{Kotani} {et~al.}(2006){Kotani}, {Kubota}, {Namiki}, {Kawai}, {Ueda},
  {Trushkin}, {Fabrika}, {Afanasiev}, {Abolmasov}, {Kinugasa}, {Nagata},
  {Irsmambetova}, {Tsukagoshi}, {Nakanishi}, {Tsuboi}, {Ozaki}, {Yanagisawa},
  {Nishiyama}, {Shimokawabe}, {Yatsu}, {Ishimura}, \& {Fujisawa}}]{kotani2006}
{Kotani}, T. et al. 2006, in VI Microquasar
  Workshop: Microquasars and Beyond

\bibitem[{{Krivonos} {et~al.}(2007){Krivonos}, {Revnivtsev}, {Churazov},
  {Sazonov}, {Grebenev}, \& {Sunyaev}}]{krivonos2007}
{Krivonos}, R., {Revnivtsev}, M., {Churazov}, E., {Sazonov}, S., {Grebenev},
  S., \& {Sunyaev}, R. 2007, \aap, 463, 957

\bibitem[{{Krivosheyev} {et~al.}(2009){Krivosheyev}, {Bisnovatyi-Kogan},
  {Cherepashchuk}, \& {Postnov}}]{krivosheyev2009}
{Krivosheyev}, Y.~M., {Bisnovatyi-Kogan}, G.~S., {Cherepashchuk}, A.~M., \&
  {Postnov}, K.~A. 2009, \mnras, 394, 1674

\bibitem[{{Kubota} {et~al.}(2007){Kubota}, {Kawai}, {Kotani}, {Ueda}, \&
  {Brinkmann}}]{kubota2007}
{Kubota}, K., {Kawai}, N., {Kotani}, T., {Ueda}, Y., \& {Brinkmann}, W. 2007,
  in Astronomical Society of the Pacific Conference Series, Vol. 362, The
  Seventh Pacific Rim Conference on Stellar Astrophysics, ed. Y.~W. {Kang},
  H.-W. {Lee}, K.-C. {Leung}, \& K.-S. {Cheng}, 121

\bibitem[{{Kubota} {et~al.}(2010){Kubota}, {Ueda}, {Fabrika}, {Medvedev},
  {Barsukova}, {Sholukhova}, \& {Goranski}}]{kubota2009}
{Kubota}, K., {Ueda}, Y., {Fabrika}, S., {Medvedev}, A., {Barsukova}, E.~A.,
  {Sholukhova}, O., \& {Goranski}, V.~P. 2010, \apj, 709, 1374

\bibitem[{{Lopez} {et~al.}(2006){Lopez}, {Marshall}, {Canizares}, {Schulz}, \&
  {Kane}}]{lopez2006}
{Lopez}, L.~A., {Marshall}, H.~L., {Canizares}, C.~R., {Schulz}, N.~S., \&
  {Kane}, J.~F. 2006, \apj, 650, 338

\bibitem[{{Magdziarz} \& {Zdziarski}(1995)}]{magdziarz1995}
{Magdziarz}, P., \& {Zdziarski}, A.~A. 1995, \mnras, 273, 837

\bibitem[{{Margon}(1984)}]{margon1984}
{Margon}, B. 1984, \araa, 22, 507

\bibitem[{{Margon} \& {Anderson}(1989)}]{margon1989}
{Margon}, B., \& {Anderson}, S.~F. 1989, ApJ, 347, 448

\bibitem[{{Margon} {et~al.}(1979){Margon}, {Grandi}, {Stone}, \&
  {Ford}}]{margon1979}
{Margon}, B., {Grandi}, S.~A., {Stone}, R.~P.~S., \& {Ford}, H.~C. 1979, \apjl,
  233, L63

\bibitem[{{Marshall} {et~al.}(2002){Marshall}, {Canizares}, \&
  {Schulz}}]{marshall2002}
{Marshall}, H.~L., {Canizares}, C.~R., \& {Schulz}, N.~S. 2002, \apj, 564, 941

\bibitem[{{Matsuoka} {et~al.}(1986){Matsuoka}, {Takano}, \&
  {Makishima}}]{matsuoka1986}
{Matsuoka}, M., {Takano}, S., \& {Makishima}, K. 1986, \mnras, 222, 605

\bibitem[{{Matt} {et~al.}(1991){Matt}, {Perola}, \& {Piro}}]{Matt1991}
{Matt}, G., {Perola}, G.~C., \& {Piro}, L. 1991, \aap, 247, 25

\bibitem[{{Medvedev} \& {Fabrika}(2009)}]{medvedev2009}
{Medvedev}, A., \& {Fabrika}, S. 2009, \mnras, 1921

\bibitem[{{Mitsuda} {et~al.}(2007){Mitsuda}, {Bautz}, {Inoue}, {Kelley},
  {Koyama}, {Kunieda}, {Makishima}, {Ogawara}, {Petre}, {Takahashi}, {Tsunemi},
  {White}, {Anabuki}, {Angelini}, {Arnaud}, {Awaki}, {Bamba}, {Boyce}, {Brown},
  {Chan}, {Cottam}, {Dotani}, {Doty}, {Ebisawa}, {Ezoe}, {Fabian}, {Figueroa},
  {Fujimoto}, {Fukazawa}, {Furusho}, {Furuzawa}, {Gendreau}, {Griffiths},
  {Haba}, {Hamaguchi}, {Harrus}, {Hasinger}, {Hatsukade}, {Hayashida}, {Henry},
  {Hiraga}, {Holt}, {Hornschemeier}, {Hughes}, {Hwang}, {Ishida}, {Ishisaki},
  {Isobe}, {Itoh}, {Iyomoto}, {Kahn}, {Kamae}, {Katagiri}, {Kataoka},
  {Katayama}, {Kawai}, {Kilbourne}, {Kinugasa}, {Kissel}, {Kitamoto}, {Kohama},
  {Kohmura}, {Kokubun}, {Kotani}, {Kotoku}, {Kubota}, {Madejski}, {Maeda},
  {Makino}, {Markowitz}, {Matsumoto}, {Matsumoto}, {Matsuoka}, {Matsushita},
  {McCammon}, {Mihara}, {Misaki}, {Miyata}, {Mizuno}, {Mori}, {Mori}, {Morii},
  {Moseley}, {Mukai}, {Murakami}, {Murakami}, {Mushotzky}, {Nagase}, {Namiki},
  {Negoro}, {Nakazawa}, {Nousek}, {Okajima}, {Ogasaka}, {Ohashi}, {Oshima},
  {Ota}, {Ozaki}, {Ozawa}, {Parmar}, {Pence}, {Porter}, {Reeves}, {Ricker},
  {Sakurai}, {Sanders}, {Senda}, {Serlemitsos}, {Shibata}, {Soong}, {Smith},
  {Suzuki}, {Szymkowiak}, {Takahashi}, {Tamagawa}, {Tamura}, {Tamura},
  {Tanaka}, {Tashiro}, {Tawara}, {Terada}, {Terashima}, {Tomida}, {Torii},
  {Tsuboi}, {Tsujimoto}, {Tsuru}, {Turner}, {Ueda}, {Ueno}, {Ueno}, {Uno},
  {Urata}, {Watanabe}, {Yamamoto}, {Yamaoka}, {Yamasaki}, {Yamashita},
  {Yamauchi}, {Yamauchi}, {Yaqoob}, {Yonetoku}, \& {Yoshida}}]{mitsuda2007}
{Mitsuda}, K. et al. 2007, \pasj, 59, 1

\bibitem[{{Namiki} {et~al.}(2003){Namiki}, {Kawai}, {Kotani}, \&
  {Makishima}}]{namiki2003}
{Namiki}, M., {Kawai}, N., {Kotani}, T., \& {Makishima}, K. 2003, \pasj, 55,
  281

\bibitem[{{Ozaki} \& {Tokimasa}(2005)}]{ozaki2005}
{Ozaki}, S., \& {Tokimasa}, N. 2005, Annual Report of the Nishi-Harima
  Astronomical Observatory (ISSN 0917-6926), No.~15, p.~15 - 29 (2005), 15, 15

\bibitem[{{Panferov} \& {Fabrika}(1993)}]{panferov1993}
{Panferov}, A.~A., \& {Fabrika}, S.~N. 1993, Astronomy Letters, 19, 41

\bibitem[{{Predehl} \& {Schmitt}(1995)}]{predehl1995}
{Predehl}, P., \& {Schmitt}, J.~H.~M.~M. 1995, \aap, 293, 889

\bibitem[{{Revnivtsev} \& {Sazonov}(2007)}]{revnivtseveandsazonov2007}
{Revnivtsev}, M., \& {Sazonov}, S. 2007, \aap, 471, 159

\bibitem[{{Takahashi} {et~al.}(2007){Takahashi}, {Abe}, {Endo}, {Endo}, {Ezoe},
  {Fukazawa}, {Hamaya}, {Hirakuri}, {Hong}, {Horii}, {Inoue}, {Isobe}, {Itoh},
  {Iyomoto}, {Kamae}, {Kasama}, {Kataoka}, {Kato}, {Kawaharada}, {Kawano},
  {Kawashima}, {Kawasoe}, {Kishishita}, {Kitaguchi}, {Kobayashi}, {Kokubun},
  {Kotoku}, {Kouda}, {Kubota}, {Kuroda}, {Madejski}, {Makishima}, {Masukawa},
  {Matsumoto}, {Mitani}, {Miyawaki}, {Mizuno}, {Mori}, {Mori}, {Murashima},
  {Murakami}, {Nakazawa}, {Niko}, {Nomachi}, {Okada}, {Ohno}, {Oonuki}, {Ota},
  {Ozawa}, {Sato}, {Shinoda}, {Sugiho}, {Suzuki}, {Taguchi}, {Takahashi},
  {Takahashi}, {Takeda}, {Tamura}, {Tamura}, {Tanaka}, {Tanihata}, {Tashiro},
  {Terada}, {Tominaga}, {Uchiyama}, {Watanabe}, {Yamaoka}, {Yanagida}, \&
  {Yonetoku}}]{takahashietal2007}
{Takahashi}, T. et al. 2007, \pasj, 59, 35

\bibitem[{{Watson} {et~al.}(1986){Watson}, {Stewart}, {King}, \&
  {Brinkmann}}]{watson1986}
{Watson}, M.~G., {Stewart}, G.~C., {King}, A.~R., \& {Brinkmann}, W. 1986,
  \mnras, 222, 261

\bibitem[{{Yuan} {et~al.}(1995){Yuan}, {Kawai}, {Brinkmann}, \&
  {Matsuoka}}]{yuan1995}
{Yuan}, W., {Kawai}, N., {Brinkmann}, W., \& {Matsuoka}, M. 1995, \aap, 297,
  451

\end{thebibliography}

\clearpage

\begin{table}
\caption{X-Ray Observation Log.}\label{table:xobslog}
\begin{center}
 \begin{tabular}{llllcl}
\hline\hline
\multicolumn{2}{c}{Start} & \multicolumn{2}{c}{End} &{Exposure}& Remark\\
 &\multicolumn{1}{c}{(MJD)} &&\multicolumn{1}{c}{(MJD)}&{(ks)} \\
\hline
\multicolumn{6}{l}{Observatory:  Suzaku}\\
2006/04/04 14:40 &(53829.6108) &2006/04/05 12:47 &(53830.5326)&38.68 &Eclipse\\
2006/04/08 11:04 &(53833.4610) &2006/04/09 10:59 &(53834.4578)&40.20 &\\
\hline
\end{tabular}
\end{center}
\end{table}

\begin{table}
\caption{Log of Optical Spectroscopic Observations}\label{table:optspeclog}
\begin{center}
\begin{tabular}{llrl}
\hline\hline
\multicolumn{2}{c}{Start} & \multicolumn{1}{c}{Exposure}& Remark\\
 &\multicolumn{1}{c}{(MJD)}&\multicolumn{1}{c}{(s)} \\
\hline
\multicolumn{4}{l}{Telescope: BTA 6 m. Observatory: SAO. PI:  S. Fabrika.}\\
2006/04/06 00:45:35 &(53831.0317) &590 &\\
\hline
\multicolumn{4}{l}{Telescope: 150 cm. Observatory:  Gunma.  PI:  K. Kinugasa.}\\
2006/03/07 19:57:47  &(53801.8297)&600(the sum of 120$\times$5)&\\
2006/04/02 18:47:00  &(53827.7826)&900(the sum of 180$\times$5)&Line not detected.\\
2006/04/05 19:15:28  &(53830.8024)&900(the sum of 180$\times$5)&\\
2006/04/06 18:39:53  &(53831.7777)&900(the sum of 180$\times$5)&\\
2006/04/07 18:40:42  &(53832.7783)&900(the sum of 180$\times$5)&\\
2006/04/09 18:38:27  &(53834.7783)&1080(the sum of 180$\times$6)&\\
2006/05/03 18:16:07  &(53858.7612)&720(the sum of 180$\times$4)&\\
2006/05/20 17:35:43  &(53875.7441)&900(the sum of 180$\times$5)&\\
\hline
\multicolumn{4}{l}{Telescope: Nayuta 2 m. Observatory:  Nishi-Harima.  PI: S. Ozaki.}\\
2006/04/03 18:04:52 &(53828.7534)&1800  &\\
2006/04/05 17:55:00 &(53830.7465) &$1800\times2$\\
2006/04/07 17:48:03 &(53832.7417) &592, 540\\
2006/04/12 08:07:16 &(53837.3384) &1800&Line not detected.\\
\hline
\multicolumn{4}{l}{Telescope: 122 cm.  Observatory: Padova-Asiago.  PI:  T. Iijima.}\\
2006/04/12 03:12:57 &(53837.1340)&1200 &\\
2006/04/14 02:37:01 &(53839.1090)&1200 &\\
\hline
 \end{tabular}
\end{center}
\end{table}

\clearpage

\begin{table}
\caption{The best fit parameters of the continuum emission
obtained by a simultaneous fit to the XIS+HXD spectra.
The uncertainties refer to statistical errors
at 90\% confidence limits for a single parameter (but 1$\sigma$ for Jet
 Temperature $T_0^{\rm line}$).}
\label{table:data}
\begin{center}
\begin{tabular}{lcccc}
\hline\hline
Parameter&\multicolumn{2}{c}{April 8--9}&\multicolumn{2}{c}{April 4--5}\\
\hline
Continuum&&&&\\
$T_{\rm 0}$ (keV)&\multicolumn{2}{c}{$27.0^{+2.1}_{-1.6}$}&\multicolumn{2}{c}{$21.1^{+1.1}_{-0.9}$}\\
R$^*$& \multicolumn{2}{c}{$1.00_{-0.41}$}&\multicolumn{2}{c}{$1.00_{-0.13}$}\\
EW$^{**}$&\multicolumn{2}{c}{$1.00^{+0.32}_{-0.38}$}&\multicolumn{2}{c}{$1.00
$ (fixed)}\\
\hline
\multicolumn{5}{l}{Line Flux: ${\rm 10^{-5}\, photon\, s^{-1}\, cm^{-2}}$}\\
\multicolumn{1}{c}{(rest-frame line energy)}&Red&Blue&Red&Blue\\
{$\rm Fe\,_{XXV}\,K\alpha \phantom{_{II}}\,(6.698\,{\rm
keV})$}&$40.5^{+1.8}_{-1.7}$&$45.8^{+2.2}_{-2.6}$&$17.90^{+0.68}_{-0.63}$&$19.61^{+0.69}_{-0.67}$\\
{$\rm Fe\,_{XXVI}\,K\alpha \phantom{_{I}}\,(6.965\,{\rm
keV})$}&$13.9^{+3.4}_{-2.9}$&$12.8^{+1.1}_{-0.85}$&$5.96\pm0.61$&$8.11^{+0.50}_{-0.49}$\\
{$\rm Ni\,_{XXVII}\,K\alpha \,(7.798\,{\rm
keV})$}&$10.8^{+1.7}_{-1.5}$&$14.8^{+3.4}_{-6.1}$&$4.95^{+0.47}_{-0.58}$&$5.04^{+0.58}_{-0.73}$\\
{$\rm Fe\,_{XXV} \,K\beta \phantom{_{II}}\,(7.897\,{\rm
keV})$}&$4.7^{+1.6}_{-1.8}$&$3.8^{+2.3}_{-1.9}$&$2.8^{+0.54}_{-1.1}$&$2.27^{+0.56}_{-0.55}$\\
{$\rm Fe\,_{XXVI} \,K\beta \phantom{_{I}}\,(8.210\,{\rm
keV})$}&$1.7^{+5.1}_{-1.8}$&$3.64 \pm 0.94$&$1.90^{+0.53}_{-0.50}$&$1.38^{+0.47}_{-0.45}$\\
{$\rm Fe\,_I \,K\alpha \phantom{_{XXVI}}\,(6.399\,{\rm
 keV})$}&\multicolumn{2}{c}{$6.9^{+2.2}_{-2.7}$}&\multicolumn{2}{c}{3.7 (fixed)}\\
\hline
Jet Temperature$^{***}$ $T_{0}^{\rm line}$
(keV)&$14.5^{+1.2}_{-1.0}$&$12.23^{+0.54}_{-0.47}$&$14.14^{+0.53}_{-0.58}$&$17.10^{+0.93}_{-0.87}$\\
\hline
Redshift&$0.0661^{+0.0008}_{-0.0009}$&$0.0064^{+0.0007}_{-0.0009}$&$0.0559^{+0.0016}_{-0.0008}$&$0.0318^{+0.0006}_{-0.0005}$\\
\hline
$\chi^2/{\rm d.o.f}$& \multicolumn{2}{c}{$810\,/\,749$}&
\multicolumn{2}{c}{$704\,/\,698$}\\
\hline
\end{tabular}
\end{center}
 {\protect\footnotesize
 $*$): $R\equiv \Omega/2\pi$, where $\Omega$ is the
  solid angle of the reflector seen from the emitter (i.e., jets).\\
 $**$): The equivalent width of an Fe$_{\rm I}$ K$\alpha$ line with respect to the
  reflection component.\\
 $***$): Determined from the line intensity ratio between
${\rm Fe\,_{XXV}\,K\alpha}$ and ${\rm Fe\,_{XXVI}\,K\alpha}$ based
on a multi-temperature jet model (see text).
 }
\end{table}

\end{document}